\documentclass[aps,groupedaddress,fleqn,twocolumn,longbibliography]{revtex4-2}
\usepackage{graphicx}
\usepackage{xcolor}
\usepackage{amsmath,amssymb}
\usepackage{float}
\usepackage{physics}
\usepackage{amsfonts}
\usepackage{verbatim}
\usepackage{flushend}
\usepackage{balance}
\usepackage{endnotes}
\usepackage{footnote}
\usepackage{adjustbox}
\usepackage{gensymb}
\usepackage{subfigure}
\usepackage{float}
\usepackage{epigraph}
\usepackage{xcolor}
\usepackage[utf8]{inputenc}
\usepackage{setspace}
\newcommand{\beq}{\begin{eqnarray}}
\newcommand{\eeq}{\end{eqnarray}}
\newcommand{\kb}{k_{\mathrm{B}}}

\newcommand{\ld}{\lambda_{\mathrm{d}}}

\newcommand{\ggrad}{\mathrm{grad}}
\usepackage{amsmath}

\usepackage[normalem]{ulem}

\begin{document}

\title{Heat transport in ionic liquids}

\author{C. Cockrell$^{1,2,*}$ \footnote{Corresponding author. E-mail: c.cockrell@bangor.ac.uk},  A. Dragović$^{3}$}
\address{$^{1,*}$ Nuclear Futures Institute, Bangor University, Bangor, LL57 1UT, UK}
\address{$^2$ School of Physical and Chemical Sciences, Queen Mary University of London, Mile End Road, London, E1 4NS, UK}
\address{$^3$  Astrophysics Group, Cavendish Laboratory, J. J. Thomson Avenue, Cambridge, CB3 0HE, UK}
\address{* Corresponding author, e-mail: c.cockrell@bangor.ac.uk}

\begin{abstract}
Heat transfer in liquids is a very challenging problem as it combines the competing effect of high frequency oscillations, which dominate liquid heat capacity, and diffusive motion, which enables mass transport and macroscopic flow. This issue is compounded by the relatively junior state of dynamical theories of liquid thermodynamics. Nevertheless, molten salts are playing an increasingly important role in industrial and energy applications and there is a pressing need to understand the mechanisms behind their irreversible transport processes. Here we use molecular dynamics simulations to investigate the heat transport of three different molten salts: LiCl, KCl, and the eutectic point of their mixture. While all simulations consider the properties of a mass element within the frame of its centre of mass, we calculate different susceptibilities which implicitly include and explicitly exclude the heat carried by partial mass currents within this frame. We find that, while the heat advected by partial mass currents in the mixture increases with increasing temperature, the heat transferred by collective vibrational motion (phonons) decreases with increasing temperature. This causes a maximum in the heat conductance with temperature in the mixtures only - in pure salts each contribution decreases monotonically with temperature. We attribute this anomaly to the extra freedom afforded to ionic motion in mixtures - in pure salts the motion of cations and anions is bound due to conservation of linear momentum. In mixtures, a coherent but diffusive collective motion is enabled by the release of Li ions from this condition by the introduction of a third species. We tentatively ascribe this coherent collective motion to the ``diffusive" phonons that have been used to explain a similar anomaly in the thermal conductivity of solids.
\end{abstract}

\maketitle

\section{Introduction}


Molten salts are poised to compose the backbone of several key green energy technologies in the coming decades \cite{Zhao2023}. In the nuclear energy sector, molten salts are proposed fuel transport fluids, coolant fluids, and pyroprocessing fluids \cite{Rosenthal1970,Merle2009,Locatelli2013,Salanne2008,Locatelli2013,Uozumi2021,Lee2011,Mitachi2022}. Furthermore, molten salts can be used as thermal storage media in order to complement renewable energy sources \cite{Carabello2021,Bhatnagar2022,Yang2010,Bauer2021}, which has advantages over conventional battery solutions in terms of resource scarcity and ecological damage. Molten salts, as ionic liquids, possess more intricate and complex microscopic dynamics than ``simple" liquids (such as elemental or molecular liquids) \cite{Hansen2003} which affect their transport properties, including thermal conductivity. The lack of a theoretical guide is a major obstacle the interpretation of experimental data and to the maturation of promising molten salt technologies \cite{Zhao2023}.

Microscopic dynamics of liquids are dominated by high-frequency ``solid-like" oscillations about quasi-equilibrium positions, with diffusive ``transits" occasionally interrupting oscillation. These transits enable the hydrodynamic behaviour of liquids but the oscillatory component of motion bestows them many properties similar to solids, as recognised and formalised by Frenkel \cite{Frenkel1955}. Recently, a phenomenological and predictive theory of liquid thermodynamics was devised on this basis, applying the Debye phonon theory of solids to collective motion in the liquid state \cite{Bolmatov2012,Proctor2020,Trachenko2016}. Liquid collective modes, like solid phonons, exhibit longitudinal and transverse polarisations, however propagating transverse phonons in liquids do not exist above a certain critical wavelength \cite{Trachenko2016}, $\ld$, called the ``dynamic length". This ``gapped" momentum state, as it has come to be known \cite{Baggioli2020}, was recently observed in experimental measurements of several liquid metals \cite{Inui2021}. The phonon theory has been expanded, predicting a universal relationship between viscosity and heat capacity \textit{via} this critical wavelength $\ld$, which was recently identified using molecular dynamics (MD) simulations in a wide range of fluids \cite{Cockrell2021,Cockrell2022,Cockrell2024b}. The phonon theory, in its elementary form, has also been expanded to qualitatively describe the evolution of thermal conductivity in liquids with a variety of bonding types \cite{Zhao2021}. These findings collectively establish ``solid-like" theories as promising approaches to understanding liquid thermal conductivity. After all, liquids support collective motion at high frequencies and wavenumbers with properties very similar to the phonons of solids. 

In solids, the increased phonon scattering with increasing temperature has a negative effect on thermal conductivity. This effect, tantamount to a decrease in the mean free path of phonons, likewise explains the decrease of the isochoric heat capacity and \textit{shear} viscosity of liquids with increasing temperature. As diffusive transits become more frequent with increasing temperature, so too does phonon scattering. On the other hand, heat (unlike shear momentum) can be transported at all wavelengths by the hydrodynamic motion enabled by diffusive transits, rendering the relationship between heat transport and collective motion less clear in liquids. Here, we perform MD simulations of molten salts, both pure and mixtures, in order to ascertain the relationship between heat transport and ionic dynamics in ionic liquids. This is facilitated by the calculation of single-particle and collective dynamical correlation functions, as well as various non-equilibrium susceptibilities related to the thermal conductivity. We find that heat transport in molten salt \textit{mixtures} shares a non-monotonic relationship, a maximum, with temperature, and that this is brought about by the collective dynamics of different ionic species as they evolve distinctly with temperature. This phenomenon is not possible in pure salts (binary liquids), unary liquids, or in solids where diffusive motion is negligible, collective or otherwise. This maximum is furthermore invisible to certain definitions of thermal conductivity which exclude the heat advected by partial or total mass currents, highlighting the rich dynamics present in complex liquids and their multifaceted effects on different thermophysical properties.

\section{Methods}

\subsection{Interatomic interactions}

The compositions we study here are LiCl, KCl, and the eutectic point of their mixture: 0.58(LiCl) 0.42(KCl). The Born-Huggins-Mayer pair potential is used to model non-bonded interactions between the ions:

\begin{equation}
    \label{eqn:bhm}
    \varphi_{IJ}(r) = A_{IJ} \exp\left(B(\sigma_{IJ} - r)\right) - \frac{C_{IJ}}{r^6} - \frac{D_{IJ}}{r^8}.
\end{equation}
Where $I$ and $J$ index species pairs. Pair paremeters for pure systems follow the Fumi and Tosi rules \cite{Fumi1964,Sangster1976}. For the eutectic (LKE), we use the mixing rules introduced by Larsen \textit{et. al.} \cite{Larsen1973} and report the parameters in Tab. \ref{tab:lkeparams}. Ions interact electrostatically and are assigned their formal charges: $z = \pm 1$.

\begin{table*}[ht]
  \centering
  \begin{tabular}{ |c|c|c|c|c|c|c| } 
  \hline
Species $i$ & Species $j$ & $A_{ij}$ (eV) & $B$ (\AA$^{-1}$) &  $\sigma_{ij}$ (\AA) & $C_{ij}$ (eV \AA${^6}$) & $D_{ij}$ (eV \AA${^6}$)  \\
 \hline
    Li & Li  & 0.4420 & 2.9455 & 1.632 & 0.04557 & 0.01875 \\ 
    Li & K & 0.3428 & 2.9455 & 2.279 & 0.7590 & 0.53057 \\
    Li & Cl & 0.291 & 2.9455 & 2.401 & 1.248 & 1.498 \\
    K & K & 0.2637 & 2.9455 & 2.926 & 15.1681  & 14.9808 \\
    K & Cl & 0.2110 & 2.9455 & 3.048 & 29.9616 & 45.5666 \\
    Cl & Cl & 0.1582 & 2.9455 & 3.170 & 73.7566 & 148.195 \\
\hline
\end{tabular}
\caption{Parameters modelling pair interactions in LKE.}
\label{tab:lkeparams}
  \end{table*}

\subsection{Molecular dynamics trajectories}  

All MD trajectories were produced using the DL\_POLY package \cite{Todorov2006}. All systems simulated in this study contain 800 atoms in total. The thermal conductivity and other parameters do not show any dependence on system size, as has been noted previously \cite{Cockrell2025}. Equilibration at the target temperatures and pressures was performed in the NPT ensemble using Nosé-Hoover thermostats and barostats, each with relaxation times of 1.0 ps, for 0.2 ns. The final states of the NPT runs are then each reseeded with 80 different velocity distributions which run in the NVE ensemble for 50 ps in order to create 80 uncorrelated initial configurations at the target density and temperature. These initial configurations are the starting points of the production runs, which are run for 1 ns in the NVE ensemble. The timestep of all simulations was 1 fs.


\subsection{Calculation of thermodynamic and transport properties}

We calculate the thermal conductivity using the Green-Kubo method \cite{Allen1991}. In single-component fluids, the thermal conductivity is identified with a single fluctuation-dissipation coefficient, $\kappa$, from the microscopic heat current autocorrelation function:
\begin{equation}
\label{eqn:gkkappa}
    \kappa \;=\; \frac{V}{3\kb T^2} \int_0^\infty \dd t \ \langle \mathbf{j}_q(0) \cdot \mathbf{j}_q(t)\rangle, 
\end{equation}
where $\mathbf{j}_q(t)$ is the microscopic heat current density:

\begin{equation}
    \label{eqn:qcurrent}
    \mathbf{j}_q(t) = \frac{1}{V} \sum_{i=1}^N \left( u^i \mathbf{v}^i + \frac{1}{2} \sum_{j\neq i} \mathbf{F}^{ij} \cdot \mathbf{v}^i \ (\mathbf{r}^{j} - \mathbf{r}^{i}) \right),
\end{equation}
with $u^i$ the total (kinetic plus potential) energy of particle $i\in 1, 2, \ldots N$, $\mathbf{F}^{ij}$ the total force impressed upon particle $i$ by particle $j$, $\mathbf{r}^{i}$ the position vector of particles $i$, and $\mathbf{v}^i$ is the velocity of particle $i$.

In multi-component fluids such as molten salts, the heat current density we defined in Eq. \ref{eqn:qcurrent} and its associated thermal conductivity in Eq. \ref{eqn:gkkappa} are not the only choices. Indeed, though in equilibrium MD simulations, $\mathbf{j}_{q}(t)$ is implicitly defined with respect to the (vanishing) barycentric velocity, the MD cell can sustain nonvanishing partial momentum densities $\mathbf{j}_I$:
\begin{equation}
    \label{eqn:partialmomentum}
    \mathbf{j}_I(t) = \frac{1}{V} \sum_i^{\{I\}} m_i \mathbf{v}_i(t),
\end{equation}
with the sum being performed only members of a single species (\textit{e.g.} Li atoms) $I$, of which there are $M$ total, such that:
\begin{equation}
    \label{eqn:momentumconservation}
    \sum_{I=1}^{M} \mathbf{j}_{I}(t) =  0,
\end{equation}
as the simulation is conducted in the centre-of-mass frame. This means that the correlation function:
\begin{equation}
    \label{eqn:correlationfunction}
    C_{q,I}(t) = \langle \mathbf{j}_q(t) \cdot \mathbf{j}_I(0) \rangle,
\end{equation}
is not trivial in multicomponent fluids. In other words, the total heat current density is coupled to mass transport in a way which isn't possible without distinct species (since the total momentum density vanishes). Macroscopically, this corresponds to the existence of local partial mass currents (partial momentum densities) in the centre of mass frame of a fluid element. The heat current density defined in Eq. \ref{eqn:qcurrent} is very convenient to calculate in MD simulations, however it does not directly relate to a useful macroscopic thermodynamic driving force. To see this, we write the phenomenological entropy source strength $\Sigma$ in an $M$-component fluid, assuming the absence of velocity gradients \cite{deGroot1984}:
\begin{equation}
    \label{eqn:entropyproduction}
    \Sigma = \mathbf{J}_{q} \cdot \ggrad\left(\frac{1}{T}\right) - \sum_{I=1}^{M-1} \mathbf{J}_I \cdot \ggrad\left({\frac{\mu_I - \mu_M}{T}}\right), 
\end{equation}
with chemical potentials $\mu_I$ of each species $I$, where $\mathbf{J}_I$ is the macroscopic version of the microscopic current densities $\mathbf{j}_I$ already defined, and having eliminated $\mathbf{J}_M$ with Eq. \ref{eqn:momentumconservation}. They are related to the thermodynamic affinities in terms of the phenomenological kinetic coefficients $L$:
\begin{equation}
    \label{eqn:Lheatflux}
    \mathbf{J}_q = L_{q,q} \ \ggrad\left(\frac{1}{T}\right) - \sum_{I=1}^{M-1} L_{q,I} \ \ggrad\left(\frac{\mu_I - \mu_M}{T}\right),
\end{equation}
\begin{equation}
    \label{eqn:Lmassflux}
    \mathbf{J}_I = L_{I,q} \ \ggrad\left(\frac{1}{T}\right) - \sum_{J=1}^{M-1} L_{I,J} \ \ggrad\left(\frac{\mu_J - \mu_M}{T}\right),
\end{equation}
where $L_{\alpha,\beta} = L_{\beta,\alpha}$ for any pair of fluxes due to Onsager's theorem of reciprocity. As $\mathbf{J}_q$ corresponds to $\mathbf{j}_q$, by standard procedure we relate the kinetic coefficient $L_{q,q}$ with the Green-Kubo integral:
\begin{equation}
    \label{eqn:LqqGK}
    L_{q,q} = \frac{V}{3 \kb} \int_0^\infty \dd t \ \langle \mathbf{j}_q(0) \cdot \mathbf{j}_q(t)\rangle,
\end{equation}
thereby relating the heat current density to temperature gradients when the terms in the sum in Eq. \ref{eqn:Lheatflux} vanish. However because temperature $T$ appears in both thermodynamic affinities, this condition requires that $T$ varies in space, but that $(\mu_I - \mu_M)/T$ remains constant, with $\mu_I$ being functions of $T$ and concentrations $n_I$ - somewhat contrived. The fluctuation-dissipation relationship of $\mathbf{j}_q$, while it does describe a susceptibility of energy flux, does not correspond to an easily measurable constitutive fluid property.

The gradients in chemical potential and temperature can be separated by expanding the derivative of $\mu$:
\begin{equation}
    \label{eqn:dmu}
    \dd\left(\frac{\mu}{T}\right) = \frac{1}{T} \dd \mu + \frac{1}{T}\left(\pdv{\mu}{T}\right)_{N,P} \dd T + \mu \ \dd\left(\frac{1}{T}\right).
\end{equation}
Noting that $\mu = \left(\pdv{G}{N}\right)_{T,P}$, with $G$ the Gibbs potential, and employing the Maxwell relation 
$$ \left(\pdv{\mu}{T}\right)_{N,P} = -\left(\pdv{S}{N}\right)_{T,P},$$
we arrive at
\begin{equation}
    \label{eqn:dmu2}
    \dd\left(\frac{\mu}{T}\right) = \frac{1}{T} \dd \mu - \frac{1}{T^2} \pdv{H}{N} \dd T,
\end{equation}
with $H = G + TS$ the enthalpy. We use this relation to rewrite Eqs. \ref{eqn:entropyproduction}-\ref{eqn:Lmassflux} in terms of separated affinities:
\begin{equation}
    \label{eqn:entropyproduction2}
    \Sigma = \mathbf{J^\prime}_{q} \cdot \ggrad\left(\frac{1}{T}\right) - \frac{1}{T}\sum_{I=1}^{M-1} \mathbf{J}_I \cdot \ggrad_{T}\left(\mu_I - \mu_M\right), 
\end{equation}
\begin{equation}
    \label{eqn:Lheatflux2}
    \mathbf{J}^{\prime}_q = L^{\prime}_{q,q} \ \ggrad\left(\frac{1}{T}\right) + \sum_{I=1}^{M-1} \frac{L^{\prime}_{q,I}}{T} \ \ggrad_{T}\left(\mu_I - \mu_M\right),
\end{equation}
\begin{equation}
    \label{eqn:Lmassflux2}
    \mathbf{J}_I = L^{\prime}_{I,q} \ \ggrad\left(\frac{1}{T}\right) + \sum_{J=1}^{M-1} \frac{L_{I,J}}{T} \ \ggrad_{T}\left(\mu_J - \mu_M\right),
\end{equation}
Here the gradients are taken at constant temperature, and the heat current density has been redefined:
\begin{equation}
    \label{eqn:heatcurrent2}
    \mathbf{J}^\prime_q = \mathbf{J}_q - \sum_{I=1}^{M-1} \mathbf{J}_I\left(h_I - h_M\right),
\end{equation}
with $h_I$ the partial specific enthalpy of species $I$. This definition of heat flux omits the advection of heat from partial mass currents in the centre of mass frame of the fluid. The kinetic coefficient $L^\prime_{q,q}$ corresponding to this flux does directly couple the flux to the temperature gradient in the absence of gradients in chemical potential, which makes it a more useful definition of thermal conductivity:
\begin{equation}
    \label{eqn:kappa2}
    \kappa^\prime = \frac{L^\prime_{q,q}}{T^2}.
\end{equation}
The corresponding microscopic heat flux, $\mathbf{j}^\prime_{q}$, however, is very challenging to calculate in MD:
\begin{equation}   
    \label{eqn:qcurrent2}
    \mathbf{j}^{\prime}_q(t) = \mathbf{j}_q(t) - \frac{1}{V} \sum_{i=1}^N \tilde{h}_i \mathbf{v}_i(t),
\end{equation}
with $\tilde{h}_i$ the partial enthalpy per particle of the species to which particle $i$ belongs. Calculations of these partial enthalpies involve the derivative of enthalpy with respect to composition, which is undesirable. Nonetheless, a comparison of Eqs. \ref{eqn:qcurrent} and \ref{eqn:qcurrent2} elucidates why $\kappa$ from Eq. \ref{eqn:gkkappa} is unhelpful macroscopically but insightful microscopically - it incorporates effects of mass transport which come about from interesting dynamics but are preferably eliminated by experimental measurements of heat conduction.

In avoiding the calculation of partial enthalpy, we can define a different thermal conductivity using Eqs. \ref{eqn:Lheatflux}-\ref{eqn:Lmassflux} and assuming a steady state  - $\mathbf{J}_I = 0$. In this case we see that
\begin{equation}
    \label{eqn:heatfluxcross}
    \mathbf{J}_q = -\frac{1}{T^2}\left(L_{q,q} - \sum_{I=1}^{M-1} \sum_{J=1}^{M-1} L_{q,I} L^{-1}_{I,J} L_{q,J} \right) \ggrad \ T,
\end{equation}
where $L^{-1}_{I,J}$ is the $I, J$ element of the inverse of the matrix whose elements are $L_{I,J}$. In the case of two and three species respectively, this yields thermal conductivities $\lambda$ of:
\begin{equation}
    \label{eqn:lambda2}
    \lambda = \frac{1}{T^2} \left( L_{q,q} - \frac{L_{q,1}^2}{L_{1,1}}\right),
\end{equation}
\begin{equation}
    \label{eqn:lambda3}
    \lambda = \frac{1}{T^2} \left( L_{q,q} - \frac{L_{q,1}^2 L_{2,2} + L_{q,2}^2 L_{1,1} - 2 L_{q,1} L_{q,2} L_{1,2}}{L_{1,1} L_{2,2} - L_{1,2}^2}\right),
\end{equation}
with $L_{q,q}$ given by Eq. \ref{eqn:LqqGK}, and the other kinetic coefficients given by similar Green-Kubo formulae:
\begin{equation}
    \label{eqn:LqIGK}
    L_{q,I} = \frac{V}{3 \kb} \int_0^\infty \dd t \ \langle \mathbf{j}_q(0) \cdot \mathbf{j}_{I}(t)\rangle,
\end{equation}
\begin{equation}
    \label{eqn:LIJGK}
    L_{I,J} = \frac{V}{3 \kb} \int_0^\infty \dd t \ \langle \mathbf{j}_{I}(0) \cdot \mathbf{j}_{J}(t)\rangle.
\end{equation}
These definitions directly relate temperature gradient to a heat flux, and correspond to the fluctuation-dissipation response of $\mathbf{j}_q(t)$ with the fluctuation-dissipation response of $\mathbf{j}_q$ and $\mathbf{j}_I$, modulated by the $\mathbf{j}_I$ and $\mathbf{j}_J$ response, subtracted. This makes $\lambda$ more closely related to $\kappa^{\prime}$ than to $\kappa$, which is seen by their close coincidence in practice \cite{Armstrong2014}. We are interested here in the transport of energy with both the exclusion and inclusion of mass transport, and note that previous studies typically calculate one of the two without reference to the other \cite{Abramova2024,Takase2011}. We therefore calculate both susceptibilities $\lambda$ and $\kappa$, which we nominate the \textit{intrinsic} and \textit{total} thermal conductivity respectively. 

In comparing the intrinsic and total thermal conductivities, their evolution with temperature, and the behaviour of the kinetic coefficients $L$ that they comprise, we gain insight into the microscopic dynamics responsible for the transfer of energy, regardless of laboratory conditions and canonical definitions.

Each quantity is calculated from each of the 80 uncorrelated trajectories at each of the target temperatures and densities. Quantities presented here are averaged over these trajectories.

\section{Results and discussion}

\begin{figure}
             \includegraphics[width=0.95\linewidth]{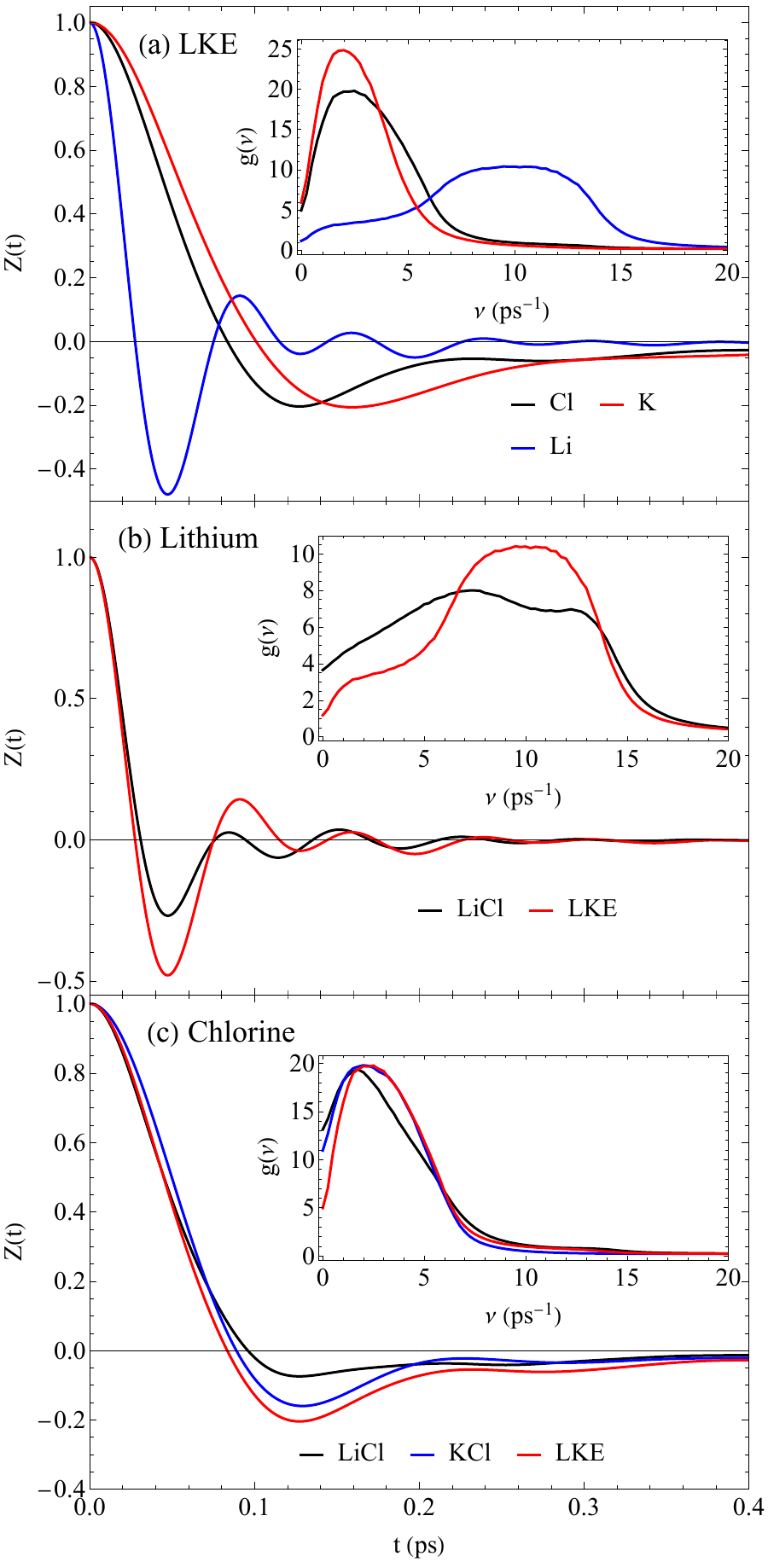}
    \caption{Velocity autocorrelation functions $Z(t)$ of (a) species in LKE at 1000 K and 1 bar; (b-c) Li and Cl in each composition at 1000 K.}
    \label{fig:vafcolumn}
\end{figure}

To better understand the microscopic dynamics facilitating heat transport, we first present the velocity autocorrelation functions (VAFs), $Z(t)$, of each species in each composition in Fig. \ref{fig:vafcolumn}. These functions are all calculated at 1 bar slightly above the melting point. This means 1000 K for the pure salts, and 700 K in the eutectic. The experimental melting point of KCl exceeds 1000 K, however the potential we use here underestimates the melting point. We choose these temperatures to be close to the melting point, and certainly far from the boiling point, of each composition. The VAF of each species in each composition shares qualitative similarities - a prominent minimum followed by a decay to zero. This first minimum, and subsequent maxima and minima where they appear, are caused by the oscillatory motion of atoms within the liquid. The rapid loss of autocorrelation, here within a picosecond or so, is caused by the diffusive component of atomic motion. High frequency vibration is disrupted by a diffusive ``jump" from one quasi-equilibrium site (``cage") to another.


The insets in Fig. \ref{fig:vafcolumn} show the density of states $g(\nu)$:
\begin{equation}
    \label{eqn:dof}
    g(\nu) = \frac{1}{\sqrt{2\pi}} \int_{0}^{\infty} \dd t \ Z(t) \exp(- 2 \pi \nu t),
\end{equation}
which expresses the frequency components of atomic motion. Diffusive motion and anharmonicity of the collective motion broaden the peaks of this function. The density of states of lithium atoms in the LKE features a broad peak at around 10 ps$^{-1}$ with a shoulder at around 3 ps$^{-1}$, the latter of which coincides with the principal peaks of the heavier ions. The VAFs and densities of states of lithium atoms and chlorine atoms in different compositions are displayed in Fig. \ref{fig:vafcolumn}b and c respectively. In pure LiCl, the peaks and troughs of $Z(t)$ are reduced compared to the mixture, indicating an increased degree of diffusive motion in the pure salt. The density of states of Li in pure LiCl exhibits a stronger component at low frequency, with the high frequency component being weaker. This is in contrast to LKE where the higher frequency component is stronger. The densities of states of chlorine atoms in different compositions do not appreciably differ, though the minimum of the VAF of LiCl is somewhat shallower than other compositions, again indicating a greater degree of diffusive motion. VAFs of potassium atoms (not displayed here for brevity) resemble those of chlorine and similarly do not significantly change with composition.

The greatest changes in motion due to changes in composition therefore pertain to the Li$^{+}$ ions. The addition of K$^{+}$ to LiCl greatly increases the participation of Li$^{+}$ in higher frequency modes. Concomitantly, the participation of Li$^{+}$ ions in low frequency modes is much diminished. In other words, the presence of two positively charged ions with different masses adds new high frequency modes to the system spectrum - this may be interpreted equivalently to the appearance of optical modes in crystalline solids when new species are added to the unit cell.

The velocity autocorrelation of lithium atoms decays to zero more quickly than that of chlorine (and potassium), though lithium generally atoms complete more oscillations in this time because they so at higher frequencies. In all liquids, diffusive jumps are infrequent enough that they may exhibit many solid-like properties over short timescales and lengthscales. This includes the ability to host transverse collective modes (phonons). As temperature increases, diffusive jumps become more frequent and atoms complete fewer oscillation periods between them. This causes a reduction in the phase space available to transverse phonons with increasing temperature. Speaking specfically, at higher temperatures the minimum wavenumber of transverse modes decreases as the frequency of diffusive jumps increases. Longitudinal phonons are not affected in this way, as a diffusive displacement can  be a component of a propagating longitudinal perturbation, whereas it must necessarily be disruptive to transverse propagation.

This reasoning explains the decrease of heat capacity and shear viscosity with increasing temperature \cite{Bolmatov2012} - the latter because it is itself a transverse property and the former due to the loss of potential energy carried by transverse phonons. These quantities always monotonically decrease with increasing temperature in liquids. The thermal conductivity, on the other hand, is a ``longitudinal property" which couples directly to density fluctuations \cite{Hansen2003}. Heat conduction can therefore propagate at all wavelengths (as in solids) and its evolution with temperature is more subtle than that of shear viscosity.

\begin{figure}
             \includegraphics[width=0.925\linewidth]{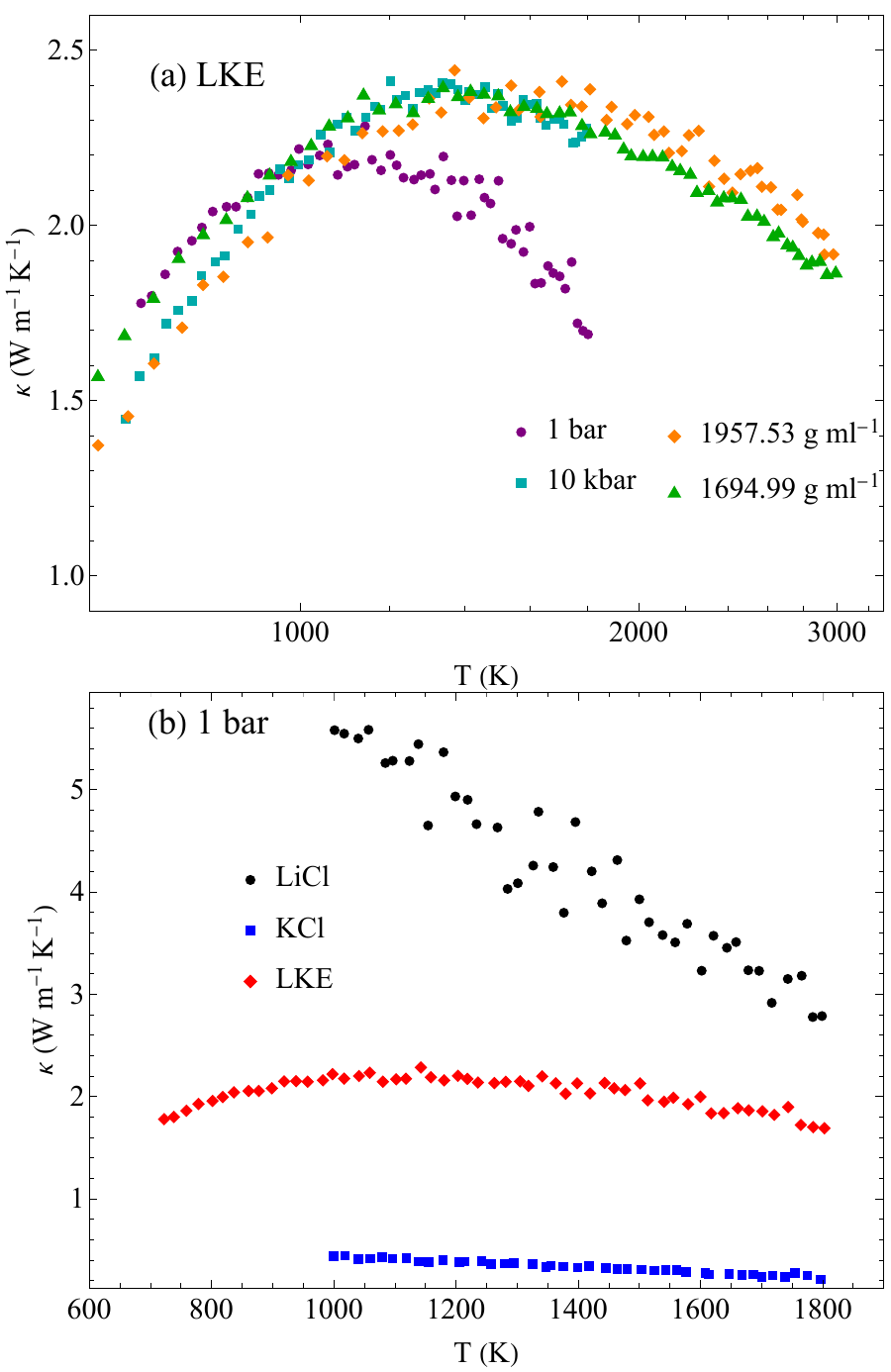}
    \caption{Total thermal conductivity $\kappa$ as a function of temperature $T$ in (a) LKE along different phase diagram paths; (b) LiCl, KCl, and LKE at 1 bar.}
    \label{fig:kappacolumn}
\end{figure}

The total thermal conductivity $\kappa$ of LKE along isobars and isochores is displayed in Fig. \ref{fig:kappacolumn}. Along each path, $\kappa$ exhibits a maximum in temperature, increasing from the melting line but decreasing again before the boiling line. This sort of response is seen in the viscosity of supercritical fluids \cite{Cockrell2021}. Furthermore, the maximum appears only in the mixture. The total thermal conductivity of the pure salts, LiCl and KCl, decreases monotonically with temperature (like the viscosity of subcritical liquids), as seen in Fig. \ref{fig:kappacolumn}b. The appearance of this maximum in the subcritical liquid mixture is therefore rather mysterious. 

\begin{figure}
             \includegraphics[width=0.9\linewidth]{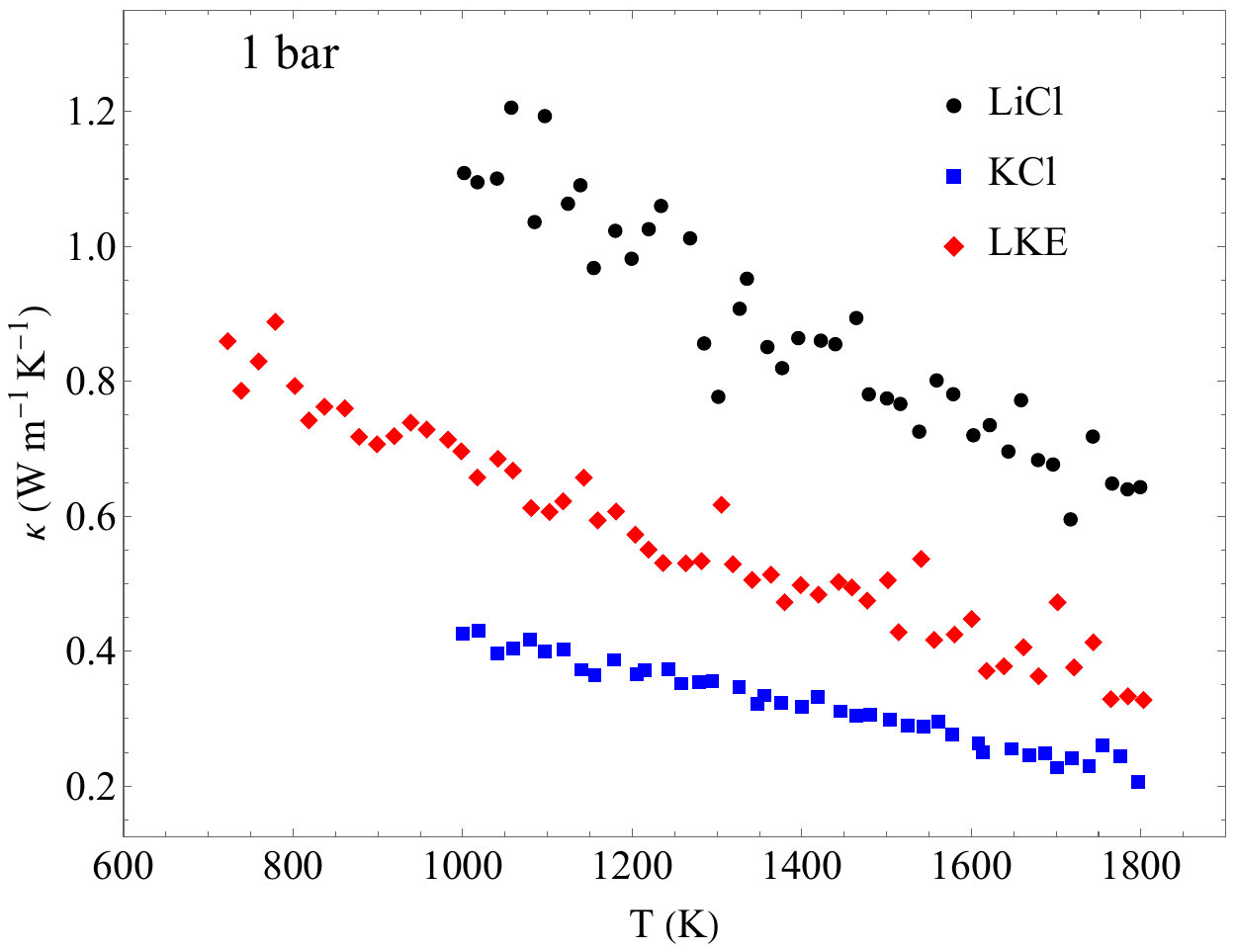}
    \caption{Intrinsic thermal conductivity $\lambda$ as a function of temperature $T$ in LiCl, KCl, and LKE at 1bar.}
    \label{fig:lambda}
\end{figure}

The intrinsic thermal conductivity $\lambda$, on the other hand, which excludes heat advected from partial momentum densities, does not present this anomaly. As depicted in Fig. \ref{fig:lambda}, $\lambda$ of each composition decreases monotonically with increasing temperature. Indeed, removal of advection terms causes a massive reduction in $\lambda$ from $\kappa$ in LiCl and LKE. On the other hand, $\lambda$ and $\kappa$ completely coincide in KCl within the error on each. The very high mobility of Li$^{+}$ ions is therefore responsible for much heat transport in LiCl versus KCl, and this advective heat transport is reponsible for the anomalous maximum in $\kappa$, though somewhat perplexingly only in mixtures.

One possible approach to the thermal conductivity of liquids is informed by the lattice thermal conductivity of solids. In electrically insulting solids, where heat is overwhelmingly carried by phonon interactions, the thermal conductivity can be approximated as
\begin{equation}
    \label{eqn:phononkappa}
    \lambda = \tilde{c}_P v l,
\end{equation}
with $\tilde{c}_P$ the isobaric heat capacity per unit volume, $v$ the speed of sound (approximating the speed of phonons) and $l$ the phonon mean free path (scattering length). In solids, outside of specific circumstances like the superionic transition, mass transport is negligible and $\lambda = \kappa$. Eq. \ref{eqn:phononkappa} results in a reduction of $\lambda$ with increasing $T$ due to the increase of scattering rate and subsequent decrease of $l$ (while the other quantities vary relatively weakly with temperature). A similar logic can be applied to liquids \cite{Cockrell2021,Cockrell2022}, with the added effect of the speed of sound noticeably decreasing with $T$ in liquids. In ``simple" liquids, therefore, thermal conductivity is expected to decrease with temperature, and for ultimately the same reasons that viscosity does. As temperature increases diffusive motion becomes more prevalent than oscillatory motion and the kinetic, ``gas-like" component of transport processes increases while the elastic, ``solid-like" component decreases, resulting in more frequent phonon scattering. 

In liquids, $\lambda$ and $\kappa$ do not coincide, and the above reasoning applies to $\lambda$ - diffusive motion, which enables partial mass currents, is responsible for the decreased mean free path, but the heat advected by the currents does not directly enter into Eq. \ref{eqn:phononkappa}. This simple picture qualitatively matches the thermal conductivity of liquids quite well \cite{Zhao2021} \textit{via} an approximation of $l$ and $v$. In pure salts, the trend of $\lambda$ with temperature is well reproduced, but the difference between compositions is not captured by this basic model \cite{Cockrell2025b}. For example, the thermal conductivity of LiCl and KCl predicted from Eq. \ref{eqn:phononkappa} are very close in magnitude.

\begin{figure}
             \includegraphics[width=0.95\linewidth]{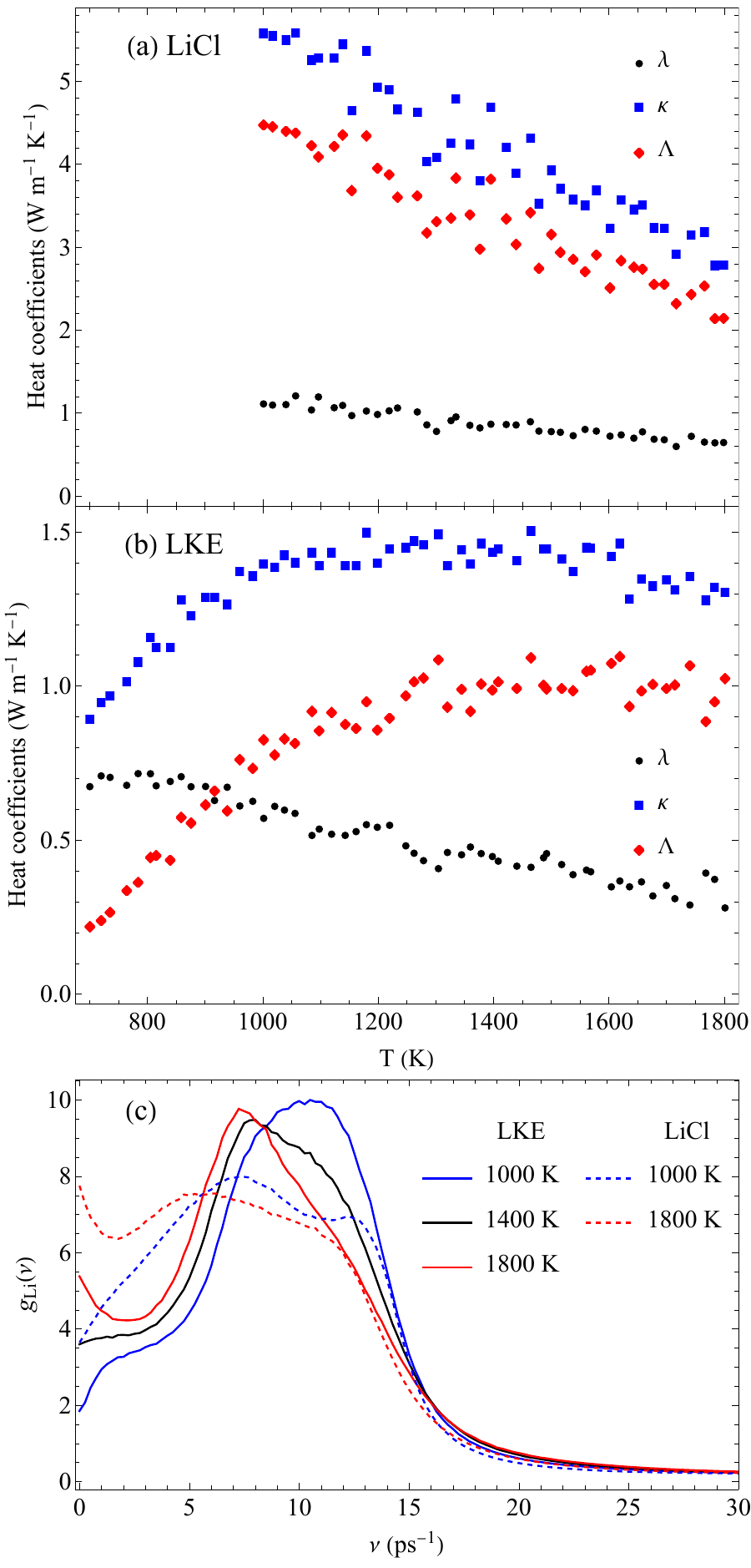}
    \caption{Heat transport coefficients $\lambda$, $\kappa$, and $\Lambda$ in (a) LiCl; and (b) LKE at 1 bar as a function of temperature; (c) Density of states of Li ions in LiCl and LKE}
    \label{fig:heatcolumn}
\end{figure}

The maxima in total thermal conductivity is therefore still difficult to explain. Atomic mobility increases and local elasticity decreases with increasing temperature. This could create an expectation that partial momentum currents and the enthalpy they advect increase with temperature - diffusive atoms escaping their local environment are faster and appear more frequently at higher temperature \cite{Cockrell2023}. This explains the increase of $\kappa$ but makes its subsequent decrease even more enigmatic. It is important to note that the increase in single-particle diffusive motion with increasing temperature does not directly correspond to the collective current densities, defined in Eq. \ref{eqn:partialmomentum}, which contribute to the fluctuation-dissipation susceptibility of the total system energy. In other words, it is \textit{a priori} conceivable that a kind of collective motion which transports heat could become more effective at over a limited temperature range. We illustrate this point by defining a new heat transport coefficient, $\Lambda$:
\begin{equation}
    \label{eqn:masslambda}
    \Lambda = \kappa - \lambda.
\end{equation}
From inspection of Eqs. \ref{eqn:lambda2} and \ref{eqn:lambda3}, we relate $\Lambda$ to the heat transported by partial mass currents. We compare $\kappa$, $\lambda$, and $\Lambda$ in LiCl and LKE in Fig. \ref{fig:heatcolumn}a-b. $\Lambda$ effectively vanishes in KCl due to the negligible mass difference \cite{Armstrong2014}. In LiCl, $\Lambda$ follows the same trend as $\kappa$ and $\lambda$ - monotonic decrease with increasing temperature. Meanwhile, in LKE, the cause of the maximum in $\kappa$ is evident as $\Lambda$ increases sharply at low temperature before plateauing at high temperature. $\Lambda$ in LKE is therefore not solely correlated with single-atom mobility, as inferred by, say, diffusion coefficients which monotonically increase with increasing temperature. From the densities of states in Fig. \ref{fig:vafcolumn}, we see that the lithium atoms are most altered by changes in composition, as discussed above. We expand upon this analysis in Fig. \ref{fig:heatcolumn}c, which shows the density of states $g_{\mathrm{Li}}(\nu)$ of lithium atoms as a function of temperature in LiCl and LKE. In both compositions, the diffusive mode centred on $\nu = 0$ becomes more prominent as temperature increases and the higher frequency modes decrease in frequency with temperature. These high-frequency peaks are, however, much sharper in LKE than in LiCl at all temperatures but especially at higher temperatures. As discussed above, the  addition of the heavier K$^{+}$ ions into LiCl strengthens the high-frequency motion of Li$^{+}$ - we see that this response is persistent at higher temperatures too. Combining this observation with the data in Figs. \ref{fig:heatcolumn}a-b, we infer the poor conduction capacity of low frequency, single-atom motion (a well-understood fact) and the presence of higher-frequency and highly-conducting collective modes, which in LKE are \textit{not} diminished with increasing temperature.

\begin{figure}
             \includegraphics[width=0.95\linewidth]{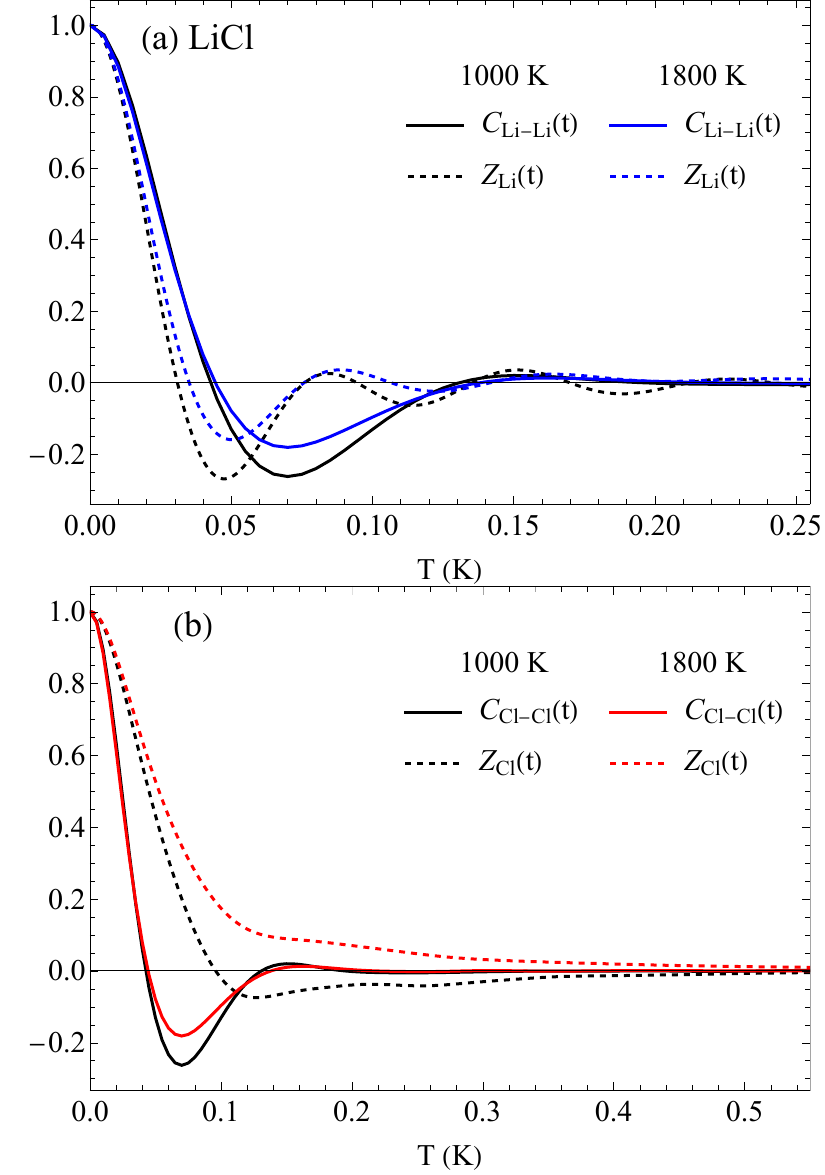}
    \caption{Comparison of current autocorrelation functions $C(t)$ and velocity autocorrelation functions $Z(t)$ for each species in LiCl.}
    \label{fig:licldynamics}
\end{figure}

\begin{figure}
             \includegraphics[width=0.95\linewidth]{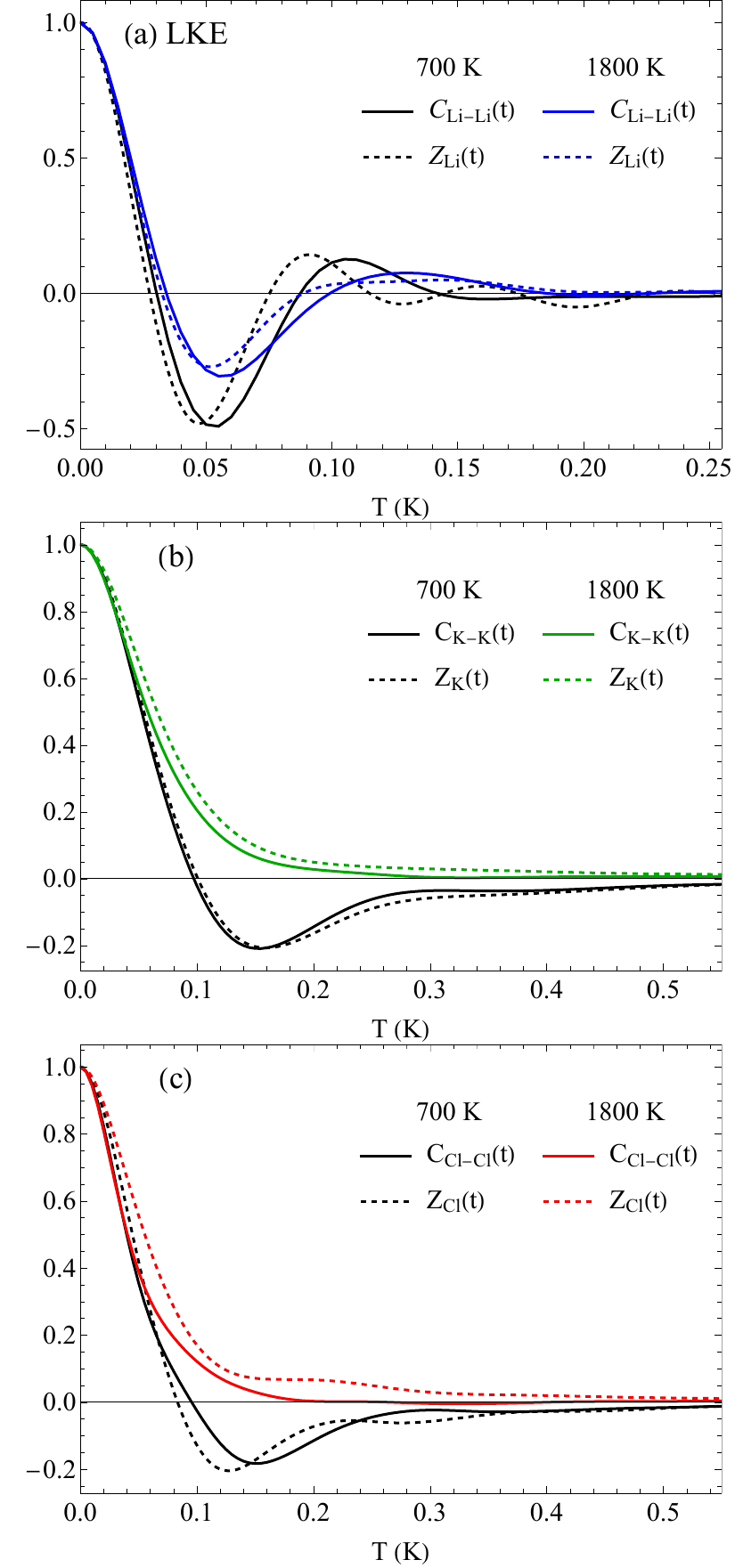}
    \caption{Comparison of current autocorrelation functions $C(t)$ and velocity autocorrelation functions $Z(t)$ for each species in LKE.}
    \label{fig:lkedynamics}
\end{figure}

We finally examine the relationship between single-particle dynamics and collective dynamics in LiCl and LKE to better understand the collective motion underlying the anomaly in $\kappa$. We define a normalised correlation function $C_{I,I}$ for each species:
\begin{equation}
    \label{eqn:cnorm}
    C_{I,I}(t) = \frac{\left\langle \mathbf{j}_{I}(t) \cdot \mathbf{j}_{I}(0) \right\rangle}{\left\langle \mathbf{j}_{I}(0) \cdot \mathbf{j}_{I}(0) \right\rangle}.
\end{equation}
This function consists of a self-part, which is proportional to the velocity autocorrelation function $Z_{I}(t)$, and cross terms, which describe the correlation between the velocities of different atoms of the same species. The extent to which $Z_{I}(t)$ and $C_{I,I}(t)$ differ is related to the coherence of the motion of species $I$. If $C$ and $Z$ exhibit different shapes and timescales, individual particles do not participate in the same vibrational modes as the collective. On the other hand, if the single-particle $Z$ resembles the collective $C$, then the motion of individual atoms closely follows collective motion, which is therefore highly coherent.

In Fig. \ref{fig:licldynamics} we plot $C_{I,I}(t)$ and $Z_{I}(t)$ for $I = $ Li, Cl at 1000 K near the melting point and 1800 K near the boiling point. Here we see that the collective current $\mathbf{j}_\mathrm{Li}$ ( $= -\mathbf{j}_\mathrm{Cl}$) autocorrelates very differently from single-atom velocities. It is easy to see that this result is inevitable in binary liquids whose masses differ. The partial momentum densities are coupled to each other by conservation of total momentum, meanwhile electrostatic forces between Li$^{+}$ and Cl$^{-}$ atoms are equal in magnitude. The collective Li$^{+}$ dynamics are obliged, by conservation of momentum, to act on a slower timescale than the single-particle dynamics, and likewise collective Cl$^{-}$ dynamics are faster than the single-particle dynamics.

The introduction of K$^{+}$ in LKE relaxes the restriction imposed by conservation of momentum. Now, the collective motion of any one species is compensated with the other two, with Li$^{+}$ and K$^{+}$ both being positively charged, and K$^{+}$ and Cl$^{-}$ both being relatively heavy. The pronounced effect of the presence of K$^{+}$ is seen in Fig. \ref{fig:lkedynamics}. Here, $C$ and $Z$ are very similar, especially for Li$^{+}$. Individual ions therefore engage in the same dynamics as the collective. The autocorrelation functions $C_\mathrm{K,K}$ and $C_\mathrm{Cl,Cl}$ decay slowly and oscillate weakly, while $C_\mathrm{Li,Li}$ oscillates quickly. Conservation of momentum is a much less restrictive condition and the collective dynamics of Li$^{+}$ atoms are much more coherent. Comparing Figs. \ref{fig:licldynamics} and \ref{fig:lkedynamics}, the collective motion of LKE contains two distinct coherent timescales - the oscillatory responses of $C_\mathrm{Li,Li}$ and $C_\mathrm{K,K}$ (the chlorine and potassium responses are roughly the same), whereas the coherent motion in LiCl only has a single timescale due to the coupling between the Li and Cl. The competition of these timescales, as they evolve differently with temperature, enables a maximum of $\kappa$ in LKE and the other mixtures. The maximum of thermal conductivity in the FLiNaK eutectic, which is lacking a theoretical explanation \cite{Robertson2022}, can be explained by the same mechanism.

We now reconcile these inferences with existing theories of conductivity in disordered matter. The increase of thermal conductivity with temperature is often called a ``glass-like" response \cite{Jaoui2023,Agne2018,Ren2024,Robertson2022}. Kittel \cite{Kittel1949} suggested the limited propagation of phonons as the culprit behind this response - the mean free path of disordered solids does not decrease as strongly with temperature as that of crystalline solids, and therefore the increase of heat capacity $c_P$ with increasing temperature causes the increase of thermal conductivity. This notion was modified by Feldman, Allen, \textit{et al.} \cite{Feldman1993a,Feldman1993b} to consider three types of collective modes in glasses: propagating modes, diffusive modes, and localised modes. At higher temperatures in glasses the diffusive collective modes become the dominant heat transfer mechanism. This rationale and others similar to it are also applicable to crystalline optical phonons \cite{Ren2024,Li2021b,Li2021,Wu2023}.

A similar mechanism is at play in liquids. Diffusive ``phonons" in solids are described as collective modes which do not propagate (in a plane wave sense) and therefore cannot be assigned a wavevector \cite{Feldman1993a,Feldman1993b}. In liquids, all collective motion is attenuated and the line between propagating and non-propagating collective motion is blurred - we comfortably ascribe wavevectors to both despite the absence of any true plane-wave motion \cite{Kryuchkov2019}. The concept is nonetheless interesting in the context of ionic liquids. The decrease of $\Lambda$ with increasing temperature in LiCl, compared with its increase in LKE, plus its eventual plateauing in LKE rule out the increase in single-atom mobility (which underlies hydrodynamic flow at vanishing wavevectors) being responsible for the maximum in $\kappa$ as (a) the frequency diffusive motion monontonically increases with increasing temperature; and (b) this response does not vary with varying system composition. The behaviour of $\kappa$ and $\Lambda$ must therefore be interpreted as the consequence of ``optic-like" diffusive collective modes whose heat transport properties are enhanced as temperature rises. The subsequent fall of $\kappa$ at high temperature is also explained in this picture - ``propagating" elastic modes are always more scattered with increasing temperature, however these ``diffusive" modes are still collective and the increase of single-atom diffusive motion likewise eventually diminishes their presence in the system spectrum. As we described above, this is enabled by the separable timescales operating in the collective currents in LKE. Similar arguments can be made to explain the evolution of heat capacity in gases and supercritical fluids due to the loss of longitudinal collective modes \cite{Trachenko2016,Cockrell2021}.

The exact nature and heat-carrying capacity of these ``diffusive" phonons is not well explored in the literature for disordered solids, and a quantitative study of them in liquids is beyond the scope of this work. Diffusive motion in liquids is very distinct from that of solids. Mass flow is inseparable from temperature response, which is why the intrinsic thermal conductivity $\lambda$ decreases with temperature - diffusive modes are coupled to mass current which are explicitly excluded from $\lambda$, an effect which does not occur in solids. The appearance of maxima in $\kappa$ in the mixtures but not in the pure salts necessitates that the collective modes enabling these maxima are complex, brought about by the combination of different oscillatory timescales due to the presence of different mass and charge combinations. 

We note that the new timescales enabling the maximum in $\kappa$ do not cause either $\kappa$ or $\lambda$ of the mixture to exceed those of LiCl. Indeed, $\lambda$ and especially $\kappa$ are far larger in pure LiCl then in the eutectic. The introduction of K$^{+}$ therefore enables a more complex response to temperature, however the highly mobile Li$^{+}$ ions which participate in high-frequency motion are less numerous in the eutectic, causing a reduction in the absolute conductivity.

\section{Conclusions}

We have calculated the the susceptibilities $\kappa$, $\lambda$, and $\Lambda$ of pure and mixed molten salts to temperature gradients using classical MD simulations. We interpret these susceptibilities as thermal conductivities with different relationships with the heat advected by partial momentum currents, only possible in fluids with distinct chemical species. Whereas the intrinsic conductivity $\lambda$ evolves monotonically with temperature - as expected by general considerations of liquid dynamics - the ``total" thermal conductivity $\kappa$ which includes the enthalpy advection term $\Lambda$ strangely exhibits a maximum in temperature.

By calculation of single-particle and collective dynamical autocorrelation functions, we explain the anomalous maximum that appears in the total thermal conductivity, $\kappa$, with the competition of multiple dynamical timescales present only in molten salt mixtures. These independent timescales are enabled by the decoupling of anion and cation mass currents - with three different chemical species, the light and fast lithium ions can access highly coherent collective dynamics without being forced to conserve the momentum of just a single other heavier species. Partial mass currents in the eutectic therefore operate on two independent timescales, whereas in pure salts the conservation of momentum permits only one. This is the qualitative difference between the dynamics of pure systems and mixtures. We suggestively ascribe coherent diffusive ``phonons", previously suggested to explain anomalies in the thermal conductivity of solids, to these extra collective degrees of freedom in the mixtures, whose contribution to heat advection grows with temperature to offset the loss of conductivity from phonon scattering (due to incoherent single-particle diffusion). The intrinsic conductivity excludes long-wavelength collective motion and is therefore sensitive only the latter effect. This explains the maximum of one type of thermal conductivity, $\kappa$, and its absence in another, $\lambda$.

We are grateful to EPSRC (grant No. EP/X011607/1). This research utilised Queen Mary's Apocrita HPC facility, supported by QMUL Research-IT. http://doi.org/10.5281/zenodo.438045, and the Sulis Tier 2 HPC platform hosted by the Scientific Computing Research Technology Platform at the University of Warwick. Sulis is funded by EPSRC Grant EP/T022108/1 and the HPC Midlands+ consortium.

\bibliography{collection_2024}

\end{document}